\documentclass{elsart}

\usepackage{natbib}

\usepackage{epsfig}

\usepackage{amssymb}

\begin{document}

\begin{frontmatter}



\title{A kinetic equation for linear fractional stable motion with applications to space plasma physics.}


\author{Nicholas W. Watkins\corauthref{cor} and Dan Credgington\thanksref{footnote2}}
\address{Physical Sciences Division, British Antarctic Survey (NERC), Cambridge, \\ CB3 0ET, UK}
\corauth[cor]{NWW is corresponding author}
\thanks[footnote2]{DC is now at London Centre for Nanotechnology, University College London, London, UK}
\ead{nww@bas.ac.uk}

\author{ Raul S\'{a}nchez}
\address{ Fusion Energy Division, Oak Ridge National Laboratory, Oak 
Ridge \\ TN 37830, USA.}
\ead{sanchezferlr@ornl.gov}

\author{Sandra C. Chapman}
\address{Centre for Fusion Space and Astrophysics, University of Warwick, Coventry \\ CV4 7AL, UK}
\ead{s.c.chapman@warwick.ac.uk}

\begin{abstract}
L\'{e}vy flights and fractional Brownian motion (fBm) have become   exemplars of  the heavy tailed jumps and long-ranged memory seen in space physics and elsewhere. Natural time series frequently combine both effects, and  Linear Fractional Stable Motion (LFSM) is a model process of this type, combining alpha-stable jumps with a memory kernel.
In contrast complex physical spatiotemporal diffusion processes where both the above effects compete-dubbed ``ambivalent"   by \citet{BrockmannEA2006}-have for many years been modelled using the fully fractional (FF) kinetic equation for the continuous time random walk (CTRW), with power laws in the pdfs of both jump size and waiting time.  We derive the analogous kinetic equation for LFSM and show that it has a diffusion coefficient with a power law in time rather than having a fractional time derivative like the CTRW.  We develop earlier comments by \citet{Lutz2001}  on how fBm differs from its fractional time process counterpart. We go on to argue more physically why LFSM and the FFCTRW might indeed be expected to differ, and discuss some preliminary results on the scaling of burst "sizes" and ``durations" in LFSM time series, with applications to modelling existing observations in space physics.
\end{abstract}


\end{frontmatter}

\parindent=0.5 cm

\section{Introduction: The need for non-Brownian models of complexity in space plasma physics.}

Fractional equations have seen at least two applications in space physics. They have been proposed in fractional kinetics-based models of microphysical processes in space plasmas \citep{MilovanovZelenyi2001}, because of the link between fractional kinetics and observed anomalous transport in magnetically confined plasmas \citep{Balescu1995,SanchezEA2005,BalescuBook}.They are also used to model magnetospheric activity, because they can describe the non-Gaussian probability density functions and long-range temporal memory seen in real time series, such as those from auroral indices and some in-situ solar wind quantities (e.g. \cite{Watkins2002}). It is not obvious, however, that it must be the {\em same} type  of fractional equation in both these contexts-i.e. reversible microphysical plasma transport equations versus equations for the evolving features of time series from a macroscopic variable. In this paper we investigate this open question, prompted in part by the stimulating suggestions made by \citet{ZaslavskyEA2007}  in response  to our earlier paper \citep{WatkinsEA2005}.

A historical preamble (section 2) will introduce the two main non-Brownian properties seen both in spatiotemporal anomalous diffusion and in many 1-dimensional time series. They are heavy tailed pdfs and long-range memory, Mandelbrot's ``Noah" and ``Joseph" effects respectively. We will also introduce some of the models in use to study them, notably the L\'{e}vy flight (ordinary L\'{e}vy motion or oLm), fractional Brownian motion (fBm) and the  continuous time random walk (CTRW). We will then discuss the highly topical problem of ``ambivalent" processes where the Noah and Joseph effects compete.

Having described the phenomena and the main types of model, in section 3 we will recap the main diffusion-like equations modifying Wiener Brownian motion (WBm) which have embodied these models in the parallel literatures on stochastic time series modelling and anomalous diffusion. While we are of course aware that concepts and models flow back and forth between these overlapping communities we feel using such a structure has made it clear how the models have developed, and exposed some surprising lacunae. 

We note that the same fractional diffusion equation holds for the pure Noah effect in both the oLm and CTRW descriptions. As noted by \citet{Lutz2001}, such an identity  no longer holds in the case of the corresponding paradigmatic models for the pure Joseph effect. We show these two different equations-for the fractional time process and fBm-and discuss why they differ. Prompted by this difference we go on to look at two corresponding self-similar models which unite the Noah and Joseph descriptions, one described by the fully fractional CTRW and one by  linear fractional stable motion (LFSM). This codification of the set of equations allows us to show that, surprisingly, a kinetic equation is ``missing" from the literature, that for LFSM. We give a simple derivation for it by direct differentiation using its well-known characteristic function.   

In section 4 we explore the application of LFSM to modelling the ``burst" sizes and durations inspired by self-organised criticality (SOC) and previously measured on magnetospheric  and solar wind time series (e.g. \citet{FreemanEA2000}).  We make simple scaling arguments building on result of \citet{KearneyMajumdar2005} to show how LFSM could  be one candidate explanation for  such ``apparent SOC" behaviour (c.f. \citet{ Watkins2002}) and compare them with numerics.  

In section 5 we discuss the interesting fact that LFSM shows additive rather than rational  ambivalent behaviour, i.e. the self-similarity exponent $H$ is an additive rather than rational expression.

In section 6 we summarise and then conclude by considering the implications of this work for the arguments made by \citet{ZaslavskyEA2007} which advocate the use of the fully fractional CTRW in space physics time series modelling.
  
\section{Time series and anomalous diffusion: Phenomena and models}

\subsection{Non-Brownian time series}

\subsubsection{Self similarity, the Hurst effect and $H$.}
 
Hurst's observation (the ``Hurst effect") of the anomalous rate of growth of range in hydrological time series, such as the height of the river Nile, was one of the first natural phenomena for which the need for a non-Brownian description was recognised. \citet{MandelbrotVanNess1968} explained the Hurst effect as being due to long range dependence in time, which they referred  to as   the ``Joseph effect". They encapsulated the Joseph effect in their seminal model, fractional Brownian motion (fBm). Like WBm, fBm has the property of self similarity under a dilation in time where $\Delta t$ goes to $\lambda \Delta t$: 
\begin{equation}
x(\lambda \Delta t) = \lambda^H x(\Delta t)
\end{equation}
 
Throughout this paper we follow \citet{EmbrechtsMaejima2002} in denoting by $H$ the self similarity exponent defined this way. Unlike WBm where $H$ is always $1/2$, in fBm $H$ takes values between 0 and 1.

\subsubsection{Long range dependence, R/s and the Joseph exponent $J$.}

To describe the growth of rescaled range ($R/s$) due to persistence,
\citet{MandelbrotVanNess1968} used the Joseph exponent $J$, where 
\begin{equation}
R/s \sim t^{J}
\end{equation}

Normal diffusion and time series modelled by random walks have $J=1/2=H$. For the particular case of fBm we still have $H=J$, and indeed the absolute value of the displacement $x$ and the mean square of displacement $<x^2>$ grow with time as $J$ and $2J$ respectively. Because of this equivalence, in the case of fBm, one can use Mandelbrot's ``R/s" method to measure $H$. More generally, however, as he emphasises \citep[e.g][p. 157]{MandelbrotRSbook} $H$ and $J$ will differ, and  R/s measures the latter.  We hence do not use the potentially confusing term ``Hurst exponent" in this paper\footnote{In the particular case of fBm we also have  that $\beta=2H+1=2J+1$, where $-\beta$ is  the exponent from the power spectral density ($S(f)\sim f^{-\beta}$)}.  

\subsubsection{Heavy tails, the Noah effect and the stability exponent $\mu$.} 

A second type of non-Brownian phenomenon had also been recognised by \citet{Mandelbrot1963}. This was the non-Gaussian 
increments, with heavy power-law tails, 
\begin{equation}
P(x) \sim x^{-(1+\mu)}
\end{equation} 
seen in financial time series and  also in many natural ones. In contrast to the Joseph effect he called this the ``Noah effect". He proposed a second paradigmatic model, ordinary Levy motion (oLm), for cases when the anomalous behaviour of the time series originates entirely from this effect, rather than long temporal memory. In oLm the index $\mu$ runs from 0 to 2. 

A self-similarity exponent $H$ remains defined for oLm, by $H=1/\mu$. However it was recognised early on that $J$ would present conceptual subtleties \citep{MandelbrotRSbook,MandelbrotWallis1969} stemming from the fact that in oLm all moments of order greater than $\mu$ are infinite, including the second order moment used in the R/s method. A finite data series drawn from an infinite variance process thus  presents  ``pseudo-Gaussian" behaviour to diagnostics such as $R/s$ or the variance. In particular $J$ takes the value it would do for an equivalent process with Gaussian amplitude distribution. $J$ is 2 for both oLm and ordinary Brownian motion, for example. As pointed out by \citet{MandelbrotWallis1969} this is an advantage if one actually just wants to measure the degree of temporal memory and not any other source of self-similarity, but it also means that R/S can longer be used to measure the full self-similarity exponent $H$ (see also the discussion in \cite{MandelbrotRSbook}). 


\subsubsection{Ambivalent time series: Noah and Joseph in competition}

Real time series usually do not exhibit just one or the other of these two limiting cases.  \citet{MandelbrotWallis1969} thus proposed that the effects modelled by fBm and oLm could be combined in a more general self similar additive model, ``fractional hyperbolic" motion, now referred to as linear fractional stable motion, LFSM, \citep[e.g.][]{EmbrechtsMaejima2002}. Mandelbrot (p.111 in \cite{ShlesingerEA1995}) described LFSM later as ``one of the ``bridges ... combining  the infinite variance feature ... and the global dependence feature ...".  He had evidently found LFSM unsatisfactory as a financial model, and remarked in 1995 that it  ``[had not] found concrete ... use". It has however, by now,  been applied to problems as diverse as communications traffic \citep{LaskinEA2002},  geophysics \citep{Painter1994}, magnetospheric physics \citep{WatkinsEA2005} and solar flares \citep{BurneckiEA2008}.  

In this selfsimilar and stable random walk there are  two contributors  
to $H$, i.e. $H=H(L,J)=L+J-1/2$. This is reminiscent of the physical finding that in nonequilibrium statistical mechanics the spatial and temporal correlation lengths need longer be the same ($\xi_{\perp} \ne \xi_{\parallel}$). As we might expect the self-similarity exponent of the Noah effect $L$ depends on the spatial exponent $\mu$ controlling the amplitude of the steps in the walk, via $L=1/\mu$. The temporal exponent $\beta$  controls the Joseph exponent $J$ for long range memory by $J=\beta/2 -1/2$. In the specific case of finite variance processes such as fBm we can equate $\beta$
to the power spectral exponent. 

\subsection{Modelling anomalous diffusion}

All the above time series models have counterparts describing the  now widely-recognised natural phenomenon of anomalous (non-Brownian) diffusion \citep{KlafterEA1996} in space and time. Particularly relevant to the space plasma case is the physics of non-equilibrium systems, and turbulent diffusion, see e.g. the reviews  of  \cite{ZaslavskyBook} and \cite{BalescuBook}. The  paradigmatic model here has been the continuous time random walk (CTRW), where the object studied is the joint probability $P(x,t)$ that a random walker makes a jump of size $x$ after waiting for a time $t$. The factorising form where $P(x,t)=P(x)P(t)$ has been particularly well explored, with power laws in $P(x)$ and $P(t)$ being used to model large jumps and long range memory effects, respectively \citep{MetzlerKlafter2000}. 

Diffusive systems in which the Noah and Joseph effects compete have recently become highly topical in complexity science \citep{BrockmannEA2006}. These authors used the CTRW with power laws in both $P(x)$ and $P(t)$ to model the space-time dynamics of dollar bills carried by travellers. They coined the phrase ``ambivalent"  diffusion for   such a ``mixed" process. \citet{WatkinsEA2005} had earlier drawn attention to the need for this type of model for space physics time series applications, but proposed the use of LFSM in such cases.  \citet{ZaslavskyEA2007} criticised their approach. While agreeing that an ambivalent model was indeed needed \citet{ZaslavskyEA2007} argued that it should be of the CTRW type and that a procedure was thus needed to define L\'{e}vy jumps in such natural time series. Very recently LFSM  has been used to model solar flare time series \citep{BurneckiEA2008}.

\section{Non-Brownian equations of motion codified}

\subsection{Random walks, the Central Limit Theorem and the diffusion equation}

The physical phenomenon of Brownian motion and its mathematical idealisation as the Wiener process are central to equilibrium statistical physics. Their links to the Central Limit Theorem (CLT), and the problem of limit distributions in general, are thus key elements in the relation of mathematics to physics. Furthermore their embodiment in diffusion equations has been essential both to practical applications and to physical understanding.
  
The familiar form of the diffusion equation is
\begin{equation}
\frac{\partial}{\partial t}P_{WBM}(x,t) = D \frac{\partial^2}{\partial x^2}P_{WBM}(x,t) \label{WBm}
\end{equation}

Parallel developments in the mathematical modelling of time series, in nonequilibrium diffusion, and turbulence, have all allowed progress in describing systems for which a description beyond Brownian motion is needed.  We will here focus mainly on the first two, and on additive models where increments are added to produce a random walk. Multiplicative models are particularly natural in turbulence but our discussion of these will be limited to the issue of multifractality in section 5.

\subsection{Heavy tailed jumps (the Noah effect)}

There are many possible ways of modifying the CLT. One involves retaining self similarity, but allowing long-range correlations in space or time. Such extensions have been studied in at least two formalisms. One is the mathematical theory of stable processes \citep{EmbrechtsMaejima2002,SamTaqBook} and other is the continuous time random walk (CTRW) \citep{MetzlerKlafter2000,ZaslavskyBook,BalescuBook}.  Both formalisms arise when the finite variance assumption of CLT is relaxed, and
 both lead to a description by a diffusion equation with a fractional derivative in space \citep{PaulBaschnagel1999,MetzlerKlafter2004}
\begin{equation}
\frac {\partial}{\partial t} P_{OLM}(x,t) = D \frac{\partial^{\mu}}{\partial x^{\mu}}P_{OLM}(x,t) \label{oLm}
\end{equation}

The two formalisms must be equivalent here because they both correspond to iid infinite variance processes (Levy flights) and so must be equivalent manifestations of the extended CLT. 

\subsection{Long ranged temporal memory (the Joseph effect)}

The situation is more subtle when the iid assumption is relaxed, rather than the finite variance one. Relaxing independence is one way to break iid.  One way to relax independence is introducing long range dependence, as studied by   \citet{MandelbrotWallis1969}. They used a self-affine process with a memory kernel, and named it fractional Brownian motion. In the CTRW formalism it was instead introduced via   a power law distribution of waiting times \citep{LindenbergWest1986} so was known as the fractional time (or temporal) process (FTP, see also \cite{Lutz2001}).  

\subsubsection{Long ranged memory in the CTRW picture: the Fractional Time Process}

The modification to the diffusion equation (\ref{WBm}) that corresponds to the FTP is \citep{Balakrishnan1986, MetzlerKlafter2000,Lutz2001}

\begin{equation}
\frac {\partial^{\beta'}}{\partial t^{\beta'} } P_{FTP}(x,t) = D \frac{\partial^{2}}{\partial x^{2}} P_{FTP}(x,t) \label{FTP}
\end{equation}
 
Note that we do not include the term describing the power law decay of the initial value here or in subsequent CTRW equations (it is retained and discussed by \citet{MetzlerKlafter2000}, see their equation 40).  
The fractional derivative in time corresponds physically to a power law in waiting time between jumps. The notation $\beta^{'}$ just indicates that this exponent need not necessarily be related to the $\beta$ we use in the following. In all the following cases $D$ is no longer the Brownian diffusion constant but simply ensures dimensional correctness in a given equation.

\subsubsection{Long ranged memory from a self-similar process: fractional Brownian motion}
Contradictory statements exist in the literature concerning the equivalent kinetic equation for fBm corresponding to equation (\ref{FTP}) for FTP. It has sometimes been asserted  \citep{ZaslavskyBook,WatkinsEA2005} that (\ref{FTP}) is also the equation of fBm. However the solution $P_{FTP}$ of  (\ref{FTP}) is now known  to be of a non-Gaussian form, given by \citep{MetzlerKlafter2000} in terms of Fox functions; whereas the pdf of fBm is by definition \citep{Mandelbrot1982,McCauley2004} Gaussian but with the standard deviation ``stretching" with time as $t^H$. The correct diffusion equation for fBm must thus, as noted by Lutz (2001), be local in time:
\begin{equation}
\frac{\partial }{\partial t} P_{FBM} = 2 H t^{2 H-1} D
\frac{\partial^{2} }{\partial x^{2}} P_{FBM} \label{fBm}
\end{equation}
given, to our knowledge, first\footnote{note the diffusion coefficient was given in \citet{Feder1988}} by   \citet{WangLung1990}. 
It can be seen by trial solution to have a solution of the required form. Further properties of the
solutions to equation (\ref{fBm}) have been given by  \citet{Lutz2001}. In addition \citet{Lutz2001} has given a corresponding fractional Langevin equation. 
fBm and the FTP are very different in the sense that, although both include temporal correlations, the kinetic equation for the former is Markovian in time and the latter is not. This is in spite of both processes being non-Markovian in respectively the Wiener and CTRW senses! This underlines (see also \citet{Lutz2001}) that these two closely related ways of modifying the CLT by introducing temporal dependence doin fact have strikingly different structures.

With hindsight we may understand why equations (\ref{fBm}) and (\ref{FTP}) differ on physical grounds. The fBm approach (\ref{fBm}) seems to us to be more macroscopic in spirit, in that a time dependent  diffusion coefficient is being imposed versus the more microscopic approach of prescribing a pdf for waiting time. However we would anticipate that the nonlinear shot noise formalism studied by \citet{EliazarKlafter2006} might allow one to derive one formalism as a limit of the other.

\subsection{Heavy-tailed jumps and long range memory together: two possible diffusion-like equations for ambivalent processes}

\subsubsection{Fully fractional CTRW and Brockmann et al's (rational) ambivalence:}

Similar questions have been asked over the years about the natural generalisation of equation (\ref{FTP}) for the fractional time process to allow for L\'{e}vy distributions of jump lengths as well as  power-law distributed waiting times. The resulting equation (corresponding in particular to Brockmann et al's ``ambivalent process") is  fractional in both  in space and time: 
\begin{equation}
\frac {\partial^{\beta'}}{\partial t^{\beta'} } P_{FFCTRW}(x,t) = D \frac{\partial^{\mu}}{\partial t^{\mu}} P_{FFCTRW}(x,t)  
\end{equation}
Again, as with the FTP  the solution for this process is known \citep{Kolokoltsov} not to be a stable (or stretched stable) distribution but rather a convolution of such distributions. 

\subsubsection{Additive ambivalence: LFSM}

As with fBm, equation (\ref{fBm}), there {\em is} a process, linear fractional stable motion (LFSM), which generalises the one represented by eqn (\ref{fBm}) to the ambivalent case. 

Its pdf $P_{LFSM}$ can be defined via its characteristic function
\begin{equation}
P_{LFSM}=\int e^{ikx} \exp (-\bar{\sigma} |k|^{\mu} t^{\mu H})  \label{LFSM}
\end{equation} 
(see for example \citep{LaskinEA2002}).

We  see that LFSM has  a  L\'{e}vy-like characteristic function: $\exp (-\bar{\sigma} |k|^{\mu} t^{\mu H})$
where the effect of  $\mu$ no longer being equal to $1/H$ is for the effective width parameter to grow like $t^{\mu H}$.
The characteristic function has the correct fBm limit, where we take $\mu$ as 2. For fBm  at any given $t$ the characteristic function is a Gaussian with width growing as $t^{2H}=t^{\beta-1}=t^{2J}$. 

That LFSM is a general stable self-affine process can be  seen by taking $k'=k\tau^H$ whereby we obtain
\begin{equation}
P_{LFSM}=t^{-H} \phi_{\mu}(x/t^H)
\end{equation}

This is a stable distribution of index $\mu$ and a prefactor which ensures the self-similarity in time discussed in section 2.1.1

The kinetic equation satisfied by (\ref{LFSM}) which generalises (\ref{fBm}) can be found by direct differentiation of (\ref{LFSM}) with respect to time to give
\begin{equation}
\frac{\partial}{\partial t} P_{LFSM} = \mu H \bar{\sigma} t^{\mu H -1} \int_{-\infty}^{\infty} e^{ikx} |k|^{\mu} \exp (-\bar{\sigma} |k|^{\mu} t^{\mu H})
\end{equation}
which, absorbing the constant $\bar{\sigma}$, and factors of $2\pi$ into $D$ can be recognised as
\begin{equation}
\frac{\partial }{\partial t} P_{LFSM} = \mu H t^{\mu H-1} D
\frac{\partial^{\mu} }{\partial x^{\mu}} P_{LFSM}
\end{equation}
using one of the standard definitions \citep[p. 110]{PaulBaschnagel1999} of a fractional derivative $\partial^{\mu}/\partial x^{\mu}$. Surprisingly equation (12) seems not have been given before in either the physics or mathematics literature.

The appropriate limits may be easily checked; in particular $\mu=2$ gives the equation of fBm found by Wang and Lung. We also remark that LFSM should be a special case of the nonlinear shot noise process studied by \cite{EliazarKlafter2006} which may allow further generalisation of the equation we have presented.  

\section{Scaling properties of ambivalent processes and their relatives}

\subsection{Burst  durations in LFSM}
\citet{Watkins2002} reviews the use of power law pdfs in ``burst" size and duration derived from time series to infer the presence of self-organised criticality (SOC) in the magnetosphere  and solar wind  (e.g. by \citet{FreemanEA2000}). Qualitatively such behaviour may also simply be an artifact of a self similar (or multifractal) time series (a possibility raised by \citet{FreemanEA2000}). To clarify this we have elsewhere (e.g. \cite{Watkins2002}) advocated the testing of SOC diagnostics using controllable self similar models. In this section we present an example (Figure 1) of the pdf of burst duration for a simulation of LFSM. LFSM was simulated using a  direct FFT-based method \citep{ChechkinGonchar2000,WatkinsEA2005}. Burst duration was defined in the standard manner as being the period for which a time series exceeds a given threshold. $\mu$ and $\beta$ were chosen as $1.5152$ and $1.58$, giving a subdiffusive $H$ value of $0.45$. These parameters are quite   typical of the geomagnetic index  $AL$ \citep{Watkins2002,WatkinsEA2005}. 

\begin{figure}
\label{figure1}
\begin{center}
\includegraphics*[width=10cm,angle=0]{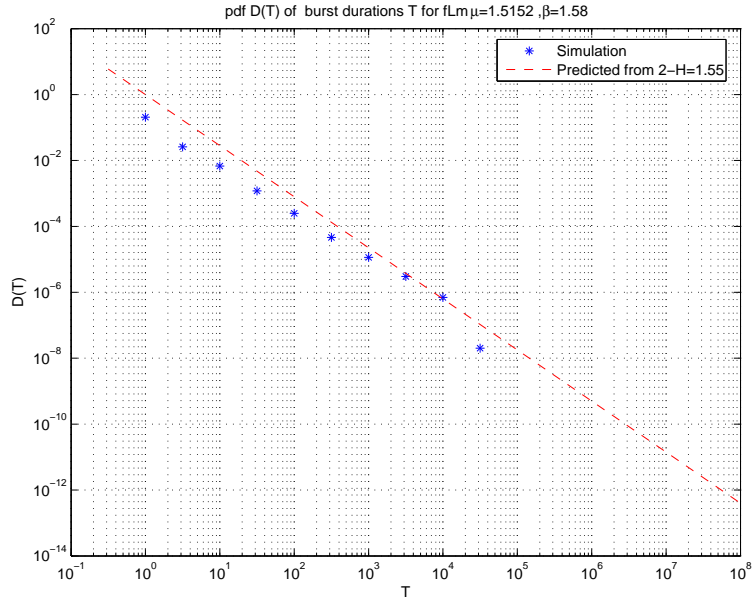}
\end{center}
\caption{Probability density $D(T)$ of a burst of duration $T$ for simulated LFSM}
\end{figure}

The expected behaviour of burst duration  for LFSM is not completely obvious from the literature. We have overlaid a power law of exponent $2-H$ as this gives the scaling of the isoset (set of threshold crossings) of a general fractal and should, we believe, thus be independent of the detailed nature of the model. This preliminary comparison suggests reasonable agreement. We should note however that detailed agreement with measured exponents is not attempted, and we would not necessarily expect it as LFSM is a very oversimplified model.

\subsection{Burst sizes in LFSM}
Figure 2 shows a representative  pdf $D(e)$ of burst size $e$, again for simulated LFSM.  Burst size was defined   as being the area above the threshold while the time series exceeds it. $\mu$ and $\beta$ were chosen as before. The power law scaling is less straightforward and there appears to be a turnover for small $e$.  

We now develop a simple scaling argument, following   \citet{KearneyMajumdar2005} to predict the scaling of the tail of the pdf of LFSM for large $e$. It is thus one candidate toy model for  such ``apparent SOC" behaviour (c.f. \citet{Watkins2002}).  \citet{KearneyMajumdar2005}  considered the zero-drift Wiener Brownian motion (WBm) case. Rather than their full analytic treatment we first recap their heuristic argument for a burst size (area) $A$ using first-passage time $t_f$. This may then be adapted to isosets and then to LFSM. They first note that for WBm $y(t)\sim t^{1/2}$ for large $t$. Then, defining $A$ by
\begin{equation}
A=\int_{t_i}^{t_f} y(t') dt'
\end{equation}
the integration implies that large $A$ scales as $t_f^{3/2}$. Simple inversion of this expression implies that $t_f$ must scale as $ t_f \sim A^{2/3}$. We independently have the standard result for first passage time for WBm:
$P(t_f) \sim t_f^{-3/2}$. To get $P(t_f)$ as a function of $A$ i.e. $P(t_f(A))$  one needs to insert the expression for $t_f$ as a function of $A$ in above equation, and in addition will need a Jacobian.  After these manipulations \citet{KearneyMajumdar2005} found that
\begin{equation}
P(A)\sim A^{-4/3}
\end{equation}

In the zero-drift but non-Brownian case we can argue that $y(t)\sim t^{H}$ for large $t$, but rather than first passage times we define an isoset-based burst size $e$ 
using the set of  up and down crossing times $\{ t_i \}$  such that the integral from
$t_i$ to $t_{i+1}$ is
\begin{equation}
e=\int_{t_i}^{t_{i+1}} y(t') dt'
\end{equation}
As  remarked by \cite{FreemanEA2000}  the points $\{ t_i \}$ form an  isoset of 
the fractal curve. For a fractal curve of self similarity exponent $H$ and dimension $D=2-H$ they have dimension $1-H$. In consequence the probability of crossings over a time interval $\tau$ goes as $\tau^{1-H}$ giving an inter-event probability scaling like $\tau^{-(1-H)}$. The pdf for inter-event intervals in the
isoset  thus scales as $\tau^{-(2-H)}$, the same scaling as for the first passage distribution in the Brownian case.

The rest of the argument goes as before. If 
$y\sim t^H$ then $e\sim t_I^{1+H}$ so $t_I \sim e^{-(1+H)}$. 
Folding in the scaling of the inter-isoset intervals ($t_I$), $P(t_I) \sim t_I^{-(2-H)}$
we can get $P(t_I)$ as a function of $e$ (i.e. $P(t_I(e))$) by inserting an expression for $t_I$ as a function of $e$ in the above equation, and again using a Jacobian. We find:
\begin{equation}
P(e) \sim e^{-2/(1+H)}
\end{equation}
which we can  check by taking the Brownian case where $H=1/2$ to retrieve $P(e) \sim e^{-4/3}$.  Comparing the numerical trial in Figure 2 with this predicted scaling exponent of $-2/(1+H)$ is again encouraging, but detailed comparison with data is postponed to a later paper.
 
\begin{figure}
\label{figure2}
\begin{center}
\includegraphics*[width=10cm,angle=0]{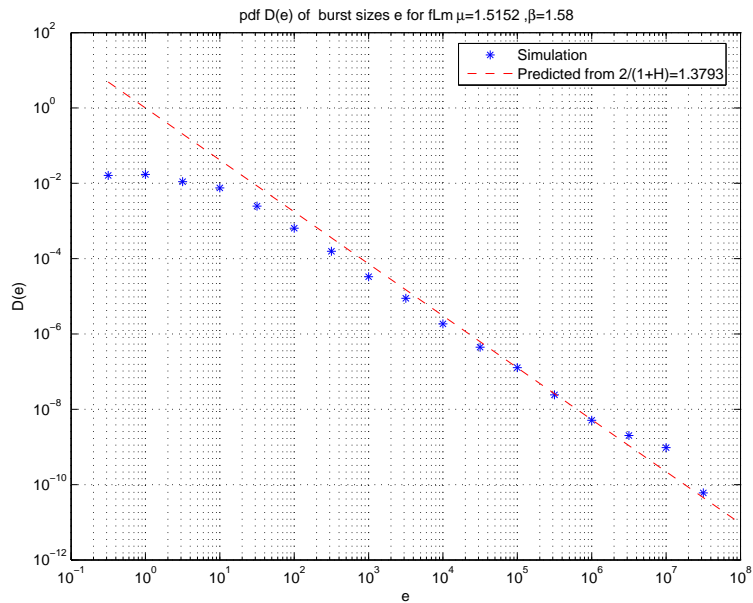}
\end{center}
\caption{Probability density  $D(e)$ of a burst of size $e$ for simulated LFSM}
\end{figure}

\section{Two sorts of ambivalence: additive and rational}
\subsection{Ambivalent processes}
Despite the primacy of mathematical descriptions, well-chosen verbal nomenclature has always been very important to physics both in guiding thought and in condensing previous insights. A recent example is the coinage by \citet{BrockmannEA2006} of the phrase ``ambivalent process". We believe that the term  should in fact be extended to {\bf all} processes in which long ranged temporal correlation and long-ranged amplitude jumps compete, not just the factorised (fully fractional CTRW) diffusive process studied by \citet{BrockmannEA2006}. Their coinage prompted us to recognise that LFSM is also an ambivalent process and we feel that this wider usage would encourage further research on the inter-relationships between these two types of ambivalent process and others. We feel it also has the merit of focusing further attention on how LFSM and the CTRW differ from L\'{e}vy Walks and multifractals.  

\subsection{Rational versus additive ambivalence}
Examining equation (12) it is clear that LFSM is also an ambivalent process  in Brockmann et al's sense, but equally clear that it differs subtly from the CTRW. The type of ambivalence previously identified by \citet{BrockmannEA2006}, comes from the factorised FF CTRW. We may rename it rational  because of clearly different scaling exponents for the  average spatial displacement $<x(t)> \sim t^{\zeta(1)}$ compared to LFSM.  Rather than $\zeta(1)$ being given by a ratio $\beta'/\mu$ of temporal and spatial exponents, as happens for the FF CTRW, for LFSM it is an additive function of ($H=J+L-1/2$) of the relevant  analogous exponents or functions thereof: 
\begin{equation}
H=[1/\mu] + [\beta/2 -1/2] - 1/2 
\end{equation} 
The difference in scaling seems  to arise because the LFSM path itself is defined as a convolution (when the long time limit is taken) between a memory kernel and a stochastic jump, and thus factorises in Fourier space.  Conversely, for the decoupled CTRW, it is the jump pdf which factorises into a temporal and spatial term. 

\subsection{A coupled space-time model: the Levy walk}

As opposed to such models which are often in some sense factorised or convolved by hand, the underlying physics itself may lead to space and time being explicitly coupled. A good example is the L\'{e}vy walk  which was introduced to deal with the problem of infinite variance by directly coupling the distribution of flight times to a long-tailed jump distribution. Well-known applications to natural systems have included the Swinney group's experiments (\cite{KlafterEA1996}) on particle tracer transport in turbulence. In a L\'{e}vy walk the exponent for $<X^2(t)>$ has several possible functional dependencies on the control parameters, a clear difference from the CTRW already remarked on by Metzler and Klafter (p. 30 of \cite{MetzlerKlafter2000}). It is of course no coincidence that early evidence for L\'{e}vy walks  came from turbulence. Space and time are unavoidably coupled in multiplicative processes, and Levy walks describe diffusion on such fields.

\subsection{Beyond self-similarity: the multifractal}

Other more complicated fractal models such as multifractals have also been extensively applied. Although only one of these four broad classes  of model may be  applicable to a specific case, availability of a wider choice of scaling exponents should  enable a better discrimination of the underlying  mechanism in at least some complex systems. 

\section{Conclusions}

In this paper we studied the question of whether one would expect the same equation to describe a time series as an anomalous diffusive process, prompted both by the stimulating suggestions of \citet{ZaslavskyEA2007} and  the highly topical work of \citet{BrockmannEA2006} on ``ambivalent" diffusion processes where the Noah and Joseph effects compete. A codification of diffusion-like equations showed that a kinetic equation was ``missing" from the literature; the one corresponding to LFSM. We gave a simple derivation for it by direct differentiation of the well-known characteristic function of LFSM.  We then made a preliminary exploration of how LFSM could model the   ``burst" sizes and durations  previously measured on magnetospheric  and solar wind time series (e.g. \citet{FreemanEA2000}).  We made simple scaling arguments building on a result of \citet{KearneyMajumdar2005} to show how LFSM could  be one candidate explanation for  such ``apparent SOC" behaviour and made preliminary comparison  with numerics. We also discussed the interesting fact that LFSM shows additive rather than rational  ambivalent behaviour, i.e. the self-similarity exponent $H$ is an additive rather than rational expression. 

All this has consequences for the interesting arguments of \citet{ZaslavskyEA2007}. We emphasise that, perhaps surprisingly, one cannot not simply equate the limiting cases (i.e. FTP and FF CTRW) of the CTRW with their fBm and LFSM analogues. This point had earlier been made for fBm by \citet{Lutz2001};  we have shown here that the problems he pinpointed must also apply to LFSM. They have interesting consequences for the applicability of the CTRW to persistent time series-derived data, leading us to believe that models which were designed with time series applications in mind, such as LFSM may be more suitable in many cases. 

Future work will include testing the above conclusions with different non-Fourier based generators for LFSM. We also plan to consider other stochastic processes, both  FARIMA and nonlinear shot noises (c.f. \citep{BurneckiEA2008}), to allow generalisation of the above initial investigations into burst size and duration. The prevalence of natural processes showing heavy tails and/or long ranged persistence suggests a relevance well beyond the initial area of application in space physics.

\section{Acknowledgements} 
 
We thank Mikko Alava, Robin Ball, Tom Chang, Aleksei Chechkin, Joern Davidsen, Mervyn Freeman, Bogdan Hnat, Mike Kearney, Khurom Kiyani, Yossi Klafter, Vassili  Kolokoltsev, Eric Lutz, Satya Majumdar and Lev Zelenyi for valuable interactions. NWW acknowledges the stimulating environment of the Newton Institute programme PDS03 in 2006 where some of this work was done. Research was carried out in part at Oak Ridge National Laboratory, managed by UT-Battelle, LLC, for U.S. DOE under contract number DE-AC05-00OR22725. SCC acknowledges support from EPSRC and STFC.

\end{document}